# Robust Time-Varying Parameters Estimation Based on I-DREM Procedure

Anton Glushchenko*, Konstantin Lastochkin*

* *V.A. Trapeznikov Institute of Control Sciences RAS, Moscow Russia (Tel: +7 495 198-17-20; e-mail: strondutt@mail.ru).*

**Abstract**: We consider a class of systems with time-varying parameters, which are written as linear regressions with bounded disturbances. The task is to estimate such parameters under the condition that the regressor is finitely exciting (FE). Considering such a problem statement, a new robust method is proposed to identify the time-varying parameters with bounded error, which could be reduced to the limit by the adjustment of such method parameters. For this purpose, the function of the system unknown parameters, which depends on time, is expanded into a Taylor series in order to turn the considered problem into the identification of the regression with piecewise-constant parameters. This results in the increase of the dimensionality of the problem to be solved. Then, the I-DREM procedure with exponential forgetting, resetting, and normalization of the regressor, which has been proposed earlier by the authors, is applied to the obtained regression. This allows one, in contrast to the known solutions, to get the dimensionality of the problem back to the initial one and provide the required exponential convergence of the parameter error to a bounded set with adjustable bound under the condition that the regressor is FE. In addition, this method guarantees that the parameter error is bounded beyond the regressor excitation interval. The above properties are proved analytically and shown via numerical simulations.

*Keywords*: finite excitation, time-varying parameters, I-DREM, identification, exponential convergence, bounded set with adjustable bound.

## 1. INTRODUCTION

Most plants could be described by linear models with unknown parameters. To provide a high-quality solution to such plants control problems, parameter identification methods are to be applied to obtain real-time estimates of the plant unknown parameters to calculate the controller parameters. At the same time, the classical identification laws (see Tao (2003)), which are widely practically applied, are based on the assumption of the unknown parameters quasi-stationarity. They are applicable only to solve the problems of the LTI plants parameter identification. Therefore, the motivation behind a number of recent studies by Hu et al. (2020), Rios et al. (2017) and others has been the development of new identification laws, which do not require the unknown parameters to be constant. The time-varying parameter identification methods proposed in the literature can be divided into direct and indirect estimation laws.

The direct estimation laws include methods that allow one to obtain values of the unknown time-varying parameters without any additional calculations. First of all, these are various versions of the direct and recursive least squares techniques with robust modifications by Mathelin and Lozano (1999), modified gradient estimation laws based on multiple filtering and averaging by Na et al. (2018). The disadvantages of these approaches are: 1) the convergence of the parameter error to a set, which bound is directly proportional to such unknown parameters change rate and 2) it is impossible to adjust the value of such bound effectively. Estimation laws with convergence over finite and fixed time Rios et al. (2017) are free of these disadvantages, but they make the identification system non-continuous. This fact may be inappropriate for some control problems.

The indirect laws of the parameter estimation include methods, in which the problem of time-varying parameters identification is transformed into the time-invariant one. In Wang et al. (2018) the method is proposed to identify not the uniformly time-varying parameters, but the piecewise-constant rate of their change. A limiting factor for this group of methods is that they are effective only if the unknown parameters change their values in accordance with the *a priori* known model, which is used to synthesize the control system. So, a detailed study of the physical nature of the parameter variation is needed. The methods of the time-varying parameters local approximation with the help of the universal polynomials Chen et al. (2011), Na et al. (2015), Zhu and Pagilla (2006) do not have such disadvantage. They allow one to represent the time-varying parameters over each regular time interval as the product of the basis functions and the constant parameters. The basis functions are proposed to be polynomials, which are obtained by a local expansion of the time-varying parameters into the Taylor series. Any classical identification loop can be applied then to estimate the locally constant parameters. The estimates of the time-varying parameters are then obtained by multiplication of the vector of the locally constant parameters estimates and the vector of the basis functions. Additionally, a special reset procedure of the identification (estimation) law is also introduced. It guarantees the continuity of the time-varying parameters estimates. In contrast to the methods from the first group, these approaches provide the convergence of the parameter error to a compact set, which bound can be adjusted by the choice of the order of the Taylor series expansion polynomial and the width of the intervals of the local expansion. But a common disadvantage of these methods is that the higher the above-mentioned polynomial order, the higher the dimensionality of the identification problem.

In general, all direct and indirect methods of time-varying parameters estimation are developed under the assumption that the necessary condition of the regressor persistent excitation (PE) or the necessary and sufficient condition of the regressor uniform injectivity is satisfied. Sastry and Bodson (2011) proved that the necessary PE condition is met if the number of spectral lines in the control action signal coincides with the number of the unknown parameters. Clearly, for many practical applications, enforcement of the PE condition through the addition of the harmonic functions to the reference signal may contradict the original control objective and lead to energy consumption increase and wear of actuators. At the same time, many real technical systems naturally meet the weaker regressor finite excitation (FE) condition.

Most of the above-considered direct and indirect identification methods, which provide the convergence of the estimation error of the time-varying parameters in the case of PE, cannot guarantee even its boundedness if this condition is not met. But the case, when the PE condition does not hold, is not taken into consideration in the course of such methods analytical derivation and further numerical experiments. So, the question of applicability of the known solutions to solve the real-world problems, for which the PE condition is often not satisfied, remains open. For example, if the PE condition is not met: 1) for many variants of the RLS estimation loop, the problem of unbounded drift of the adaptive gain arises (various modifications have been proposed to solve it, including the method of Hu et al. (2020) to check online whether the PE condition holds); 2) considering methods on the basis of the Taylor series expansion of the time-varying parameters, the one faces unrestricted growth of the estimates of the unknown parameters due to the incorrect interpolation at the moments of the estimation loop reset.

Thus, the aim of the current research is to develop an identification (estimation) loop that, first, ensures exponential convergence of the parameter error of time-varying parameters identification into a bounded set over the regressor finite excitation interval, and, second, keep the parameter error bounded beyond the excitation time range.

To achieve the stated goals, in this research it is proposed to approximate the current "true" value of the time-varying parameters by their value for the previous regular time moment, i.e. to use the method of the first-order interpolation. Such approximation allows one to transform the problem of identification of the arbitrary time-varying parameters to the one of the piecewise-constant parameters over the regular time intervals. Interpolation of the first order is a particular case of the Taylor series expansion, and for its implementation no more than the first two summands of such a series is needed. This will increase the dimensionality of the identification problem by no more than a factor of two. We propose to use a modified I-DREM procedure with interval-based filters with exponential forgetting to identify piecewise-constant parameters over regular time intervals. Thus, the exponential convergence of the piecewise-constant parameter estimates to the neighborhood of their true values at each regular time interval will be provided. In addition, I-DREM will allow us to reduce the dimensionality of the identification problem back to the original (initial) one. The boundedness of the piecewise-constant parameter errors over the regular intervals will be shown to be sufficient for the boundedness of the parameter error of the time-varying parameters when the regressor is finitely exciting.

The paper is organized as follows. The rigorous problem statement is in Section 2. The main result is presented in Section 3. The results of the numerical simulations are shown in Section 4.

*1.1 Preliminaries*

The following notation will be used in this paper: $L_\infty$ is the space of essentially bounded functions, $\|.\|$ is the Euclidean norm of a vector, *floor*(.) is the operation of rounding values to integers, $I_{n \times n}$ is the identity matrix, $O_{n \times 1}$ is the unit vector of length *n*, $0_{n \times n}$ is the zero matrix, $\lambda_{min}(.)$, $\lambda_{max}(.)$ are the minimum and maximum eigenvalues respectively, *f(t)* is the function, which depends on time (the argument *t* may be further omitted for brevity). Also, the following definitions will be used in this research.

**Definition 1.** *The regressor $\varphi(t) \in L_\infty$ is finitely and continuously exciting ($\varphi \in$c-FE) over the interval $\left[t_r^+; t_e\right]$ if there exist $0 \leq t_r^+ < t_e$, $\alpha_1 > \alpha_2 > 0$ and $T_s > 0$ such that*:

$$\int_{t_r^+}^{t_e} \varphi(\tau)\varphi^T(\tau)d\tau \geq \alpha_1 I,$$

$$\forall t \in \left[t_r^+; t_e\right) \int_{t}^{t+T_s} \varphi(\tau)\varphi^T(\tau)d\tau \geq \alpha_2 I, \quad (1)$$

*where I is the identity matrix, $\alpha_1$, $\alpha_2$ are levels of excitation.*

**Definition 2.** *The Taylor series expansion of $p^{th}$ order of the function f(t) in the neighborhood of the point t = a is the following power series*:

$$f(t) = f(a) + \frac{t-a}{1!}f^{\langle 1 \rangle}(a) + \ldots + \frac{(t-a)^p}{p!}f^{\langle p \rangle}(a) +$$

$$+ \int_a^t \frac{(t-\zeta)^{p+1}}{(p+1)!} f^{\langle p+1 \rangle}(\zeta)d\zeta, \quad (2)$$

*where $f^{\langle p \rangle}(a)$ is the $p^{th}$ derivative of the function $f(t)$.*

## 2. PROBLEM STATEMENT

Let the linear regression with time-varying parameters be considered:

$$y(t) = \Theta^T(t)\omega(t) + d(t), \quad (3)$$

where $y \in R^{1 \times m}$, $\omega \in R^{n \times m}$ are the measurable function and regressor, $\Theta \in R^n$ is the vector of the unknown time-varying parameters, $d \in R^{1 \times m}$ is the unknown disturbance, $m \leq n$. For the sake of brevity, the dependence of these variables on time will be further omitted. Considering the system (3), the following assumptions are made.

***Assumption 1.*** *The regressor* $\omega$ *is bounded* $\|\omega\| \leq \omega_{max}$.

***Assumption 2.*** *The parameters* $\Theta$ *and their* $p+1$, $p = \overline{0,1}$ *derivatives are bounded:* $\|\Theta\| \leq \Theta_{max}$ *and* $\|\Theta^{\langle p+1 \rangle}\| \leq \Theta_{max}^{\langle p+1 \rangle}$.

***Assumption 3.*** *The disturbance* $d$ *is bounded* $|d| \leq d_{max}$.

Assumption 1 could be hold by multiplication of the regression (3) by the normalization function $n_s(t) = \frac{1}{1+\omega(t)\omega^T(t)}$. Assumptions 2 and 3 are classical, as far as the problem of the time-varying parameters identification is concerned. The estimates $\omega_{max}, \Theta_{max}, \Theta_{max}^{\langle p+1 \rangle}$ are considered to be unknown. They are needed only to prove theorems/propositions, but not to apply the procedure being proposed. A great number of the parameter identification problems can be reduced to model (3) using different parametrization methods. And in such cases, the disturbance $d$ usually depends on the rate of the unknown parameters change, measurement noise, initial conditions of the applied filters and different external factors of stochastic nature.

The aim is to derive an algorithm, which: (**Goal 1**) $\forall t \in [t_r^+; t_e]$ is able to identify the time-varying parameters $\Theta$ with bounded error, which can be reduced to its limit with the help of such algorithm parameters adjustment; (**Goal 2**) provides the parameters error boundedness $\forall t > t_e$.

### 3. MAIN RESULT

*3.1 Taylor-series-based approximation*

Let the increasing time sequence be introduced:
$$t_i = T \, floor\left(\frac{t - t_r^+}{T}\right), \; i \in \mathbb{N}. \quad (4)$$

where $T_s < T \ll t_e - t_r^+$ is the time window, over which each certain Taylor series expansion is made.

Following Definition 2, let the continuous function $\Theta(t)$ be expanded into Taylor series in the neighborhood $T$ of the point $t_i$:

$$\forall t \in [t_i; t_{i+1}): \Theta(t) = \overbrace{\Theta_i + (t-t_i)\dot{\Theta}_i}^{\delta_0(t)} + \underbrace{\int_{t_i}^{t} \frac{(t-\zeta)^2}{2} \ddot{\Theta}(\zeta)d\zeta}_{\delta_1(t)}, \quad (5)$$

where $\Theta_i = \Theta(t_i), \dot{\Theta}_i = \dot{\Theta}(t_i)$ are the approximations of the time-varying parameters and the rate of their change over the regular time interval $[t_i; t_{i+1})$, $\|\delta_1\| \leq 0{,}5\ddot{\Theta}_{max}T^2$ is the bounded reminder of the first order ($p = 1$) Taylor series expansion, $\|\delta_0\| \leq \dot{\Theta}_{max}T$ is the bounded reminder of the zero-order ($p = 0$) Taylor series expansion.

Considering that $\Theta_i$ and $\dot{\Theta}_i$ change their values at time points $t_i$ like stepwise functions, the piecewise-constant approximation of the piecewise-continuous function $\Theta(t)$ $\forall t \geq 0$ is written as:

$$\Theta(t) = \underbrace{\sum_{i=0}^{\infty} \Delta_1(t_i)h(t-t_i)}_{\Theta_i} + (t-t_i)\underbrace{\sum_{i=0}^{\infty} \Delta_2(t_i)h(t-t_i)}_{\dot{\Theta}_i} + \delta_1, \quad (6)$$

where $\|\Delta_1(t_i)\| = \|\Theta_i - \Theta_{i-1}\| \leq \Delta_{1max}$ is the parameters $\Theta_{i-1}$ change at time point $t_i$, $\|\Delta_2(t_i)\| \leq \Delta_{2max}$ is the parameters $\dot{\Theta}_{i-1}$ change at the time point $t_i$, $h(t-t_i)$ is the unit step function at the time point $t_i$.

Let the following notation be introduced into (6):
$$\Theta(t) = \Lambda(t,t_i)\theta_i + \delta_1,$$
$$\Lambda(t,t_i) = \begin{bmatrix} L(t,t_i) & & \\ & \ddots & \\ & & L(t,t_i) \end{bmatrix} \in R^{n \times 2n}, \quad (7)$$
$$L(t,t_i) = \begin{bmatrix} 1 & t-t_i \end{bmatrix}, \; \theta_i = \begin{bmatrix} \Theta_i & \dot{\Theta}_i \end{bmatrix}^T \in R^{2n},$$

Then (7) is substituted into (3):
$$y = \theta_i^T \Lambda^T(t,t_i)\omega + \delta_1^T\omega + d = \theta_i^T\overline{\omega} + \varepsilon, \quad (8)$$
where $\overline{\omega} = \Lambda^T(t,t_i)\omega \in R^{2n \times m}$, $\|\varepsilon\| \leq 0{,}5\ddot{\Theta}_{max}T^2\omega_{max} + d_{max}$.

***Assumption 4.*** *The continuous finite excitation of the regressor is preserved after the transformation of* $\omega$ *into* $\overline{\omega}$ *in (8), i.e.* $\omega \in c\text{-}FE \Leftrightarrow \overline{\omega} \in c\text{-}FE$.

Thus, using the Taylor series expansion, the regression (3) with time-varying parameters $\Theta$ is transformed into the regression (8) with piecewise-constant parameters $\theta_i$, but the dimensionality of the identification problem is $2n$ now instead of the original $n$.

*3.2 I-DREM procedure*

In order to keep the original dimensionality of the identification problem, the procedure I-DREM is applied (Glushchenko et al. (2021)), according to which the regression (8) is filtered with an interval-based integral filter with exponential forgetting:

$$y_f = \omega_f\theta_i + \varepsilon_f, y_f(t_i) = \varepsilon_f = 0_{2n}, \; \omega_f(t_i) = 0_{2n \times 2n}$$
$$\omega_f^T = \int_{t_i}^{t} e^{-\int_0^{\tau}\beta d\tau_1}\overline{\omega}(\tau)\overline{\omega}^T(\tau)d\tau; \; y_f^T(t) = \int_{t_i}^{t} e^{-\int_0^{\tau}\beta d\tau_1} y(\tau)\overline{\omega}^T(\tau)d\tau, \quad (9)$$
$$\varepsilon_f^T(t) = \int_{t_i}^{t} e^{-\int_0^{\tau}\beta d\tau_1}\varepsilon(\tau)\overline{\omega}^T(\tau)d\tau$$

where $y_f \in R^{2n}$, $\omega_f \in R^{2n \times 2n}$, $\beta > 0$ is the forgetting factor. Following Aranovskiy et al. (2016), the regression (9) is multiplied by the adjunct matrix $adj\{\omega_f\}$, and the property $adj\{\omega_f\}\omega_f = det\{\omega_f\}I_{2n \times 2n}$ is used to obtain:

$$\overline{\Upsilon} = \overline{\Omega}\theta_i + \overline{\mu},$$
$$\overline{\Upsilon} := adj\{\omega_f\}y_f, \overline{\Omega} := det\{\omega_f\}, \overline{\mu} := adj\{\omega_f\}\varepsilon_f. \quad (10)$$

where $\overline{\Omega} \in R$ is the scalar regressor, $\overline{\Upsilon} \in R^{2n}$, $\overline{\mu} \in R^{2n}$.

The properties of the regression (10) are described in the following proposition.

**Proposition 1.** *If* $\omega \in$ *c-FE over the time range* $[t_r^+; t_e]$ *and Assumptions 1-4 hold, then:*

1) $\forall t \geq 0 \ \overline{\Omega}(t) \geq 0$ *and* $\forall t_a \geq t_b, \ \overline{\Omega}(t_a) \geq \overline{\Omega}(t_b);$

2) $\exists T_{0k} \in [t_k; t_{k+1}), \forall t \in [T_{0k}; t_{k+1}) \ 0 < \overline{\Omega}(T_{0k}) \leq \overline{\Omega}(t) \leq \Omega_{UB};$

*where* $t_k = T \, floor\left(\frac{t-t_r^+}{T}\right)$, $k = \overline{0, k_{max}}$, $k_{max} = floor\left(\frac{t_e - t_r^+}{T}\right)$.

3) $\forall t \in [t_i; t_{i+1}) \ \|\overline{\mu}\| \leq \mu_{max}$.

*Proof of Proposition 1 is postponed to Appendix A.*

As the regressor $\overline{\Omega}$ is scalar according to (10), then the dimensionality of the identification problem under consideration can be reduced back to $n$ through exclusion from (10) equations, which correspond to $\dot{\Theta}_i$:

$$\Upsilon = \overline{\Omega}\Theta_i + \mu,$$
$$\Upsilon := H\overline{\Upsilon}, \mu := H\overline{\mu}, \quad (11)$$

where $\Upsilon \in R^n$, $\mu \in R^n$, $H = [I_{n \times n}, \ 0_{n \times n}]$ is the transition matrix.

Let the estimation law of the piecewise-constant approximation $\Theta_i$ be introduced:

$$\dot{\hat{\Theta}} = \begin{cases} -\frac{\gamma_0}{\overline{\Omega}^2}\overline{\Omega}(\overline{\Omega}\hat{\Theta} - \Upsilon), & \text{if } \overline{\Omega} \geq \kappa \\ -\Gamma\omega[\hat{\Theta}^T\omega - y]^T - \sigma\Gamma\hat{\Theta}, & \text{otherwise} \end{cases}, \quad (12)$$

where $\gamma_0 > 0$, $\sigma > 0$, $\Gamma > 0_{n \times n}$, $\kappa \in \left(0; \min_k\{\overline{\Omega}(T_{0k})\}\right)$.

**Remark 1.** *The division by $\overline{\Omega}^2$, used in (12), is "safe" operation as, when* $\omega \in$ *c-FE, the regressor* $\overline{\Omega} > 0 \ \forall t \in [T_{0k}; t_{k+1})$.

Based on the identification (estimation) law (12), let the notion of the piecewise-constant parameters $\Theta_i$ identification error $\tilde{\Theta}_i$ be introduced. So the time-varying parameters $\Theta$ identification error is defined as:

$$\tilde{\Theta}_i = \hat{\Theta}(t_i + T) - \Theta_i,$$
$$\tilde{\Theta} = \hat{\Theta}(t_i + T) - \Theta = \hat{\Theta}(t_i + T) - \Theta_i - (t - t_i)\dot{\Theta}_i - \delta_1 = \tilde{\Theta}_i - \delta_0. \quad (13)$$

Here the identification error of the first approximation of $\Theta_i$ is calculated using the last estimate of $\hat{\Theta}$ over $[t_i; t_{i+1})$.

The introduced notations from (13) and the one for the scalar case ($n = m = 1$) regressor are illustrated in Fig. 1.

In terms of parameter errors (13), the properties of the estimation law (12) are summarized in the following Theorem.

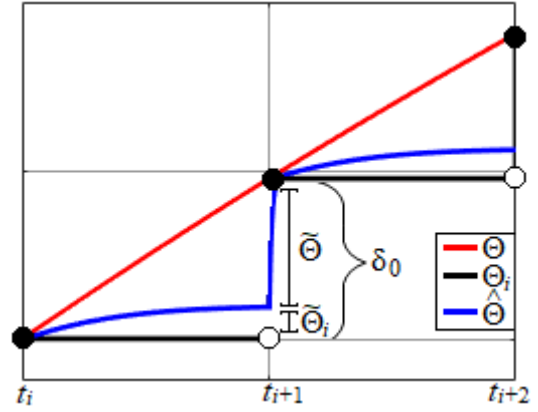

Figure 1. Parameter errors notation.

**Theorem 1.** *If* $\omega \in$ *c-FE over the time range* $[t_r^+; t_e]$ *and assumptions 1-4 hold, then the law (12) guarantees the following properties:*

(a) $\forall i \cap k$ *the error* $\tilde{\Theta}$ *is bounded as follows*:

$$\|\tilde{\Theta}\| \leq a^{k+1}e^{-(k+1)\gamma_0\Delta T}\|\tilde{\Theta}_i(t_r^+)\| + \left(\frac{a^k e^{-\gamma_0 k\Delta T}-1}{ae^{-\gamma_0\Delta T}-1}\right)ae^{-\gamma_0\Delta T}\Delta_{1max} + \\ + \left(\frac{a^{k+1}e^{-(k+1)\gamma_0\Delta T}-1}{ae^{-\gamma_0\Delta T}-1}\right)b + \dot{\Theta}_{max}T, \quad (14)$$

*Moreover, if* $k \to \infty$ *and* $\gamma_0 > \frac{1}{\Delta T}ln\left(\frac{1}{a}\right)$, *then*

$$\|\tilde{\Theta}\| \leq \frac{1}{1-ae^{-\gamma_0\Delta T}}\left(ae^{-\gamma_0\Delta T}\Delta_{1max}+b\right)+\dot{\Theta}_{max}T, \quad (15)$$

*and the upper bound in* (15) *could be reduced if: 1) the parameter T value is decreased, 2) the parameter $\gamma_0$ or the multiplication $\beta^2\overline{\Omega}(T_{0k})$ value is increased (by change of $\beta$ value).*

(b) $\forall i \not\cap k$ *the error* $\tilde{\Theta}$ *is bounded in accordance with*:

$$\|\tilde{\Theta}\| \leq a_1^{i+1}e^{-(i+1)\eta T}\|\tilde{\Theta}_i(t_e)\| + \left(\frac{a^i e^{-\eta i T}-1}{ae^{-\eta T}-1}\right)a_1 e^{-\eta T}\Delta_{1max} + \\ + \left(\frac{a^{i+1}e^{-(i+1)\eta T}-1}{ae^{-\eta T}-1}\right)b_1 + \dot{\Theta}_{max}T. \quad (16)$$

*The proof of Theorem and definitions of* $a$, $a_1$, $b$, $b_1$, $\eta$, $\Delta T$ *are presented in Appendix B.*

So, the proposed estimation law (12) provides the achievement of both **Goals** stated in Section 2. In contrast to the known solutions Chen et al. (2011), Na et al. (2015), Zhu and Pagilla (2006), in which the Taylor series expansion is applied, the proposed algorithm makes it possible to keep the initial dimension $n$ of the identification problem and to provide the boundedness of the parameter error $\forall t > t_e$.

## 4. NUMERICAL EXAMPLE

Let the regression (3) be defined as follows for the numerical experiments:

$$\omega = \begin{bmatrix} 3\sin(4\pi t) \\ \begin{cases} 2,5, \ t < 10 \\ 2,5\sin(4\pi t), \ t \geq 10 \end{cases} \end{bmatrix}; \ \Theta = \begin{bmatrix} 2+\sin(t) \\ 3+\cos(0,5t) \end{bmatrix}. \quad (17)$$

It is easy to check that the chosen regressor satisfies condition (1) when $t_r^+ = 0$, $t_e = 10$, $\alpha_1 = 45$ and, for an instance, $T_s = 0,1$; $\alpha_2 = 0,345$. When $t \geq 10$, the condition (1) is not met due to linear dependence of the regressor elements. The parameters of (4), (9) and (12) are defined as follows:

$$T = 0,25; \beta = \frac{0,05}{T}; \hat{\theta}(0) = [0\ 0]^T; \gamma_0 = 100; \kappa = 10^{-9}; \quad (18)$$
$$\Gamma = 0,75 I_{2\times 2}; \sigma = 10^{-4}.$$

Two experiments have been conducted to demonstrate the properties of the proposed algorithm. The first of them was to consider the case when $d = 0$. The disturbance was defined as the function with bias $d = rand(1) - 0,5$ for the second experiment. The obtained results for both of them are shown in Fig.2.

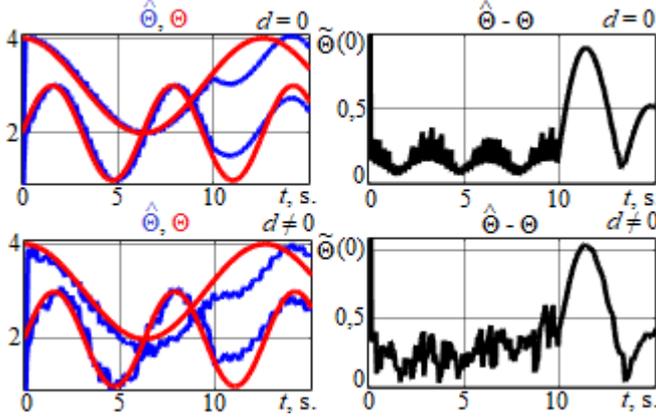

Figure 2. Experiments results.

The transients shown in Fig. 2 demonstrate that the proposed identification algorithm functioned correctly, and also confirm the achievement of both stated **Goals** and theoretical conclusions drawn in the proof of *Theorem* 1.

## 5. CONCLUSION

The time-varying parameters identification algorithm proposed in this research allows one to avoid a number of drawbacks of the known methods, which are also based on the expansion of the unknown parameters into the Taylor series. Namely, in comparison with them, it does not increase the dimensionality of the identification problem and guarantees the parameter error boundedness when the convergence condition is not satisfied (beyond the boundaries of the continuous finite excitation time interval). Considering practical applications, the obtained results can be useful for various composite identification laws, namely to improve the quality of the unknown time-varying parameters estimation.

**Appendix A. Proof of Proposition 1**

According to (9), $\text{sign}\left(\exp\left(-\int_0^\tau \beta d\tau_1\right)\overline{\omega}(\tau)\overline{\omega}^T(\tau)\right) \geq 0_{2n\times 2n}$.

As a result, $\dot{\omega}_f \geq 0_{2n\times 2n}, \omega_f \geq 0_{2n\times 2n}$ and $\Omega = det\{\omega_f\} \geq 0\ \forall t \geq 0$. Then, $\dot{\Omega} = det\{\dot{\omega}_f\} \geq 0$ holds and $\Omega(t_a) \geq \Omega(t_b)\ \forall t_a \geq t_b$. This completes the proof of the first part of the proposition.

To prove its second part, let the bounds on the damped multiplier from the definition of $\omega_f$ be obtained:

$$\forall t \in [t_i; t_{i+1}): e^{-\beta T} \leq e^{-\int_0^t \beta d\tau_1} \leq 1. \quad (A1)$$

As $\omega \in$ c-FE, then, considering (1), (A1) and following the mean value theorem and the equality $det\{cI_{n \times n}\} = c^n \ \forall c$, the lower bound on $\Omega(t_{k+1})$ is obtained:

$$\Omega(t_{k+1}) = det\left\{\int_{t_k}^{t_{k+1}} e^{-\int_0^\tau \beta d\tau_1} \overline{\omega}(\tau)\overline{\omega}^T(\tau) d\tau\right\} \geq e^{-n\beta T}\alpha_2^n. \quad (A2)$$

As $\Omega(t_a) \geq \Omega(t_b) \ \forall t_a \geq t_b$, and $\Omega(t_{k+1}) \geq e^{-n\beta T}\alpha_2^n$, then $\exists T_{0k} \in [t_k; t_{k+1})$ so as $\Omega(T_{0k}) = e^{-n\beta T}\alpha_2^n$ and $\forall t \in [T_{0k}; t_{k+1}]$ holds:

$$\Omega(t) \geq \Omega(T_{0k}) > 0. \quad (A3)$$

To obtain the upper bound on the regressor $\Omega$ over the time range $[t_k; t_{k+1}]$, the definitions (8), (9) are taken into consideration and the mean value theorem is applied:

$$\omega_f(t) = \int_{t_k}^{t_{k+1}} e^{-\int_0^\tau \beta d\tau_1} \overline{\omega}(\tau)\overline{\omega}^T(\tau) d\tau \leq \frac{\delta_k}{\beta} I_{2n \times 2n},$$
$$\delta_k = \underset{t_k \leq t \leq t_{k+1}}{\text{ess sup}} \lambda_{max}\left(\overline{\omega}(\tau)\overline{\omega}^T(\tau)\right). \quad (A4)$$

Then, considering $det\{cI_{n \times n}\} = c^n \ \forall c$, $\forall t \in [t_k; t_{k+1}]$ the following inequality holds:

$$\Omega(t) \leq \left(\delta_k \beta^{-1}\right)^n = \Omega_{UB}. \quad (A5)$$

Having combined (A3) and (A5), the proof of the second part of the proposition is completed.

To prove its third part, the upper bound of the disturbance $\varepsilon_f^T(t)$ from (8) $\forall t \in [t_i; t_{i+1}]$ is obtained:

$$\varepsilon_f^T(t) \leq \int_{t_i}^{t_{i+1}} e^{-\int_0^\tau \beta d\tau_1} \varepsilon(\tau)\overline{\omega}^T(\tau) d\tau \leq \frac{\left(0{,}5T^2\ddot{\Theta}_{max}\omega_{max}^2 + d_{max}\omega_{max}\right)\Lambda_{max}}{\beta} O_{1 \times 2n}. \quad (A6)$$

Applying $adj\{cI_{n \times n}\} = c \cdot adj\{I_{n \times n}\} \ \forall c$, it is obtained from (A4), (A6) that:

$$\|\overline{\mu}\| \leq \overline{\mu}_{max} = \frac{\left(0{,}5T^2\ddot{\Theta}_{max}\omega_{max}^2 + d_{max}\omega_{max}\right)\delta_k \Lambda_{max}}{\beta^2}\sqrt{2n}, \quad (A7)$$

as was to be proved in the third part of the proposition.

**Appendix B. Proof of Theorem 1**

To prove part (a) of Theorem, let the function $V_1 = \frac{1}{2}(\hat{\Theta} - \Theta_i)^T \Gamma^{-1}(\hat{\Theta} - \Theta_i)$ be considered. The derivative of this function has an upper bound over the time range $[t_i; T_{0k})$ as $\Omega < \kappa$:

$$\dot{V}_1 \leq \frac{1}{2}\left(-\sigma\|\hat{\Theta} - \Theta_i\|^2 + \left(d_{max} + T\dot{\Theta}_{max}\omega_{max}\right)^2 + \sigma\Theta_{max}^2\right). \quad (B1)$$

Then, solving (B1), the bound on $\hat{\Theta}(T_{0k}) - \Theta_i$ is obtained:

$$\|\hat{\Theta}(T_{0k}) - \Theta_i\| \leq a_1 e^{-\eta(t - t_i)}\|\hat{\Theta}(t_i) - \Theta_i\| + b_1,$$
$$a_1 = \sqrt{\frac{\lambda_{max}(\Gamma^{-1})}{\lambda_{min}(\Gamma^{-1})}}, \ \eta = \frac{\sigma}{2\lambda_{max}(\Gamma^{-1})}, \ b_1 = a_1 \frac{\left(d_{max} + T\dot{\Theta}_{max}\omega_{max}\right) + \sqrt{\sigma}\Theta_{max}}{\sqrt{\sigma}} \quad (B2)$$

Let the time range $[T_{0k}; t_i + T)$ be considered and the function $V_2 = \frac{1}{2}(\hat{\Theta} - \Theta_i)^T(\hat{\Theta} - \Theta_i)$ be introduced. Considering $\kappa < \underset{k}{min}\{\Omega(T_{0k})\}$, $\Omega \geq \kappa$ and *Proposition* 1, $\forall i \cap k$ its derivative has upper bound $\forall t \in [T_{0k}; t_i + T)$:

$$\dot{V}_2 \leq -\frac{1}{2}\gamma_0\|\hat{\Theta} - \Theta_i\|^2 + \frac{1}{2}\frac{\gamma_0\mu_{max}^2}{\Omega^2(T_{0k})}. \quad (B3)$$

Then, solving (B3), the bound on $\hat{\Theta}(t_i + T) - \Theta_i$ is obtained:

$$\|\hat{\Theta}(t_i + T) - \Theta_i\| \leq e^{-\frac{1}{2}\gamma_0(t_i + T - T_{0k})}\|\hat{\Theta}(T_{0k}) - \Theta_i\| + \frac{\mu_{max}}{\Omega(T_{0k})}. \quad (B4)$$

Substituting (B2) into (B4), the upper bound of the parameter error of the $\Theta_i$ approximation at the time point $t_i + T$ is written:

$$\|\hat{\Theta}(t_i + T) - \Theta_i\| \leq ae^{-\gamma_0 \Delta T}\|\hat{\Theta}(t_i) - \Theta_i\| + b,$$
$$a = a_1 e^{-\eta(T_{0k} - t_i)}, \ \Delta T = \frac{t_i + T - T_{0k}}{2}, \ b = e^{-\gamma_0 \Delta T}b_1 + \frac{\mu_{max}}{\Omega(T_{0k})}. \quad (B5)$$

Using the summation formula for geometric series, the equation $\|\hat{\Theta}(t_i) - \Theta_i\| = \|\hat{\Theta}(t_{i-1} + T) - \Theta_{i-1} + \Theta_{i-1} - \Theta_i\|$ and applying the equation (B5) recursively $k$ times (in other words, till the time point $t_r^+$), it is obtained:

$$\|\tilde{\Theta}_i\| \leq a^{k+1}e^{-(k+1)\gamma_0 \Delta T}\|\tilde{\Theta}(t_r^+)\| + \left(\frac{a^k e^{-\gamma_0 k \Delta T} - 1}{ae^{-\gamma_0 \Delta T} - 1}\right)ae^{-\gamma_0 \Delta T}\Delta_{1max}$$
$$+ \left(\frac{a^{k+1}e^{-(k+1)\gamma_0 \Delta T} - 1}{ae^{-\gamma_0 \Delta T} - 1}\right)b. \quad (B6)$$

The upper bound on the parameter error $\tilde{\Theta}$ is written as:

$$\|\tilde{\Theta}\| \leq \|\tilde{\Theta}_i - \delta_0\| \leq \|\tilde{\Theta}_i\| + \|\delta_0\| \leq \|\tilde{\Theta}_i\| + \dot{\Theta}_{max}T. \quad (B7)$$

Substituting (B6) into (B7), the equation (14) is obtained. As $\gamma_0 > \frac{1}{\Delta T}ln\left(\frac{1}{a}\right)$, then, considering $\gamma_0 > 0$, $ae^{-\gamma_0 \Delta T} < 1$ holds. Then the limit $\underset{k \to \infty}{lim}\left(ae^{-\gamma_0 \Delta T}\right)^k = 0$ holds. Considering that, the equation (14) is transformed into (15). From that, as the following limits hold:

$$\underset{\gamma_0 \to \infty}{lim} ae^{-\gamma_0 \Delta T} = 0, \ \underset{T \to 0}{lim}\dot{\Theta}_{max}T = 0, \ \underset{\substack{\beta^2 \Omega_{LB} \to \infty \\ \gamma_0 \to \infty}}{lim} b = 0, \quad (B8)$$

follows that the upper bound in (15) can be reduced. This completes the proof of part (a) of *Theorem* 1.

To prove the part (b), it is taken into consideration that $i \not\cap k \ \forall t > t_e$ and $\Omega \leq det\left\{\int_{t_i}^{t} \overline{\omega}(\tau)\overline{\omega}^T(\tau) d\tau\right\} = 0$ holds. So, as $\Omega \geq 0$ in accordance with *Proposition* 1, it is concluded that $\Omega = 0$ and $\Omega < \kappa$. Then, solving (B1), $\forall i \not\cap k$ the upper bound of the parameter error of $\Theta_i$ at time point $t_i + T$ is written as:

$$\|\hat{\Theta}(t_i + T) - \Theta_i\| \leq a_1 e^{-\eta T}\|\hat{\Theta}(t_i) - \Theta_i\| + b_1. \quad (B9)$$

Then, by analogy with (B6) and (B7), the upper bound of the parameter error from (16) $\forall i \not\cap k$ is obtained. This completes the proof of part (b) of *Theorem* 1.